\documentclass[fleqn,10pt]{wlscirep}
\title{Contagion on complex networks with persuasion}

\author[1]{Wei-Min Huang}
\author[2,3]{Li-Jie Zhang}
\author[1,3,*]{Xin-Jian Xu}
\author[1,3]{Xinchu Fu}
\affil[1]{Department of Mathematics, Shanghai University, Shanghai 200444, China}
\affil[2]{Department of Physics, Shanghai University, Shanghai 200444, China}
\affil[3]{Institute of Systems Science, Shanghai University, Shanghai 200444, China}

\affil[*]{xinjxu@shu.edu.cn}

\begin{abstract}
The threshold model has been widely adopted as a classic model for studying contagion processes on social networks. We consider asymmetric individual interactions in social networks and introduce a persuasion mechanism into the threshold model. Specifically, we study a combination of adoption and persuasion in cascading processes on complex networks. It is found that with the introduction of the persuasion mechanism, the system may become more vulnerable to global cascades, and the effects of persuasion tend to be more significant in heterogeneous networks than those in homogeneous networks: a comparison between heterogeneous and homogeneous networks shows that under weak persuasion, heterogeneous networks tend to be more robust against random shocks than homogeneous networks; whereas under strong persuasion, homogeneous networks are more stable. Finally, we study the effects of adoption and persuasion threshold heterogeneity on systemic stability. Though both heterogeneities give rise to global cascades, the adoption heterogeneity has an overwhelmingly stronger impact than the persuasion heterogeneity when the network connectivity is sufficiently dense.
\end{abstract}
\begin{document}

\flushbottom
\maketitle
%
%
\thispagestyle{empty}


\section*{Introduction}

In a population of interacting agents, a small fraction of individuals holding an opinion opposite to the one held by the majority may trigger large cascades under certain circumstances. Examples include the diffusion of cultural fads~\cite{SB1992}, the outbreak of political unrest~\cite{SL1994}, and the spread of rumor~\cite{YM2004}, etc. All these processes can be studied by contagion models~\cite{DJW2002,CM2007}, in which inactive (or susceptible) individuals are activated (or infected) by contacts with active neighbours. Different from mathematical models of biological contagions~\cite{KR2007} where each susceptible-infective contact results in infection spreading with an independent probability, social contagions are not just the spread from one specific source to another. In many situations, the possibility of an individual being activated depends on who are active among her social contacts; and this is particularly true for people who interact with each other in a social network~\cite{PG2014}. Therefore, the individual heterogeneity and the interaction structure are two key factors in determining whether a cascade occurs or not. One of the prototypes for studying such contagion dynamics is the threshold model originated from the seminal work of Schelling~\cite{TCS1973} on residential segregation, and subsequently developed by Granovetter~\cite{MG1978} in the study on social influences. According to the general definition of the threshold model, an individual adopts a new opinion only if a critical fraction (the Watts model~\cite{DJW2002}) or number (the Centola-Macy model~\cite{CM2007}) of her friends have already been activated. This required fraction/number of adopters in the neighbourhood is defined as the {\it threshold}. We call it adoption threshold hereafter.

Theoretical studies on complex contagions have mainly been developed along two distinct directions. On the one hand, researchers have been focusing on large ensembles of network systems, where interactions among individuals are described by a complex network. In 2002, Watts~\cite{DJW2002} studied the threshold model on Poisson and power-law networks. It was found that having heterogeneous nodal degrees helps enhancing systemic stability compared to that of homogeneous networks with a Poisson degree distribution. Threshold heterogeneity, however, has an opposite effect. By adopting an analytical approach on locally tree-like networks, Gleeson and Cahalane~\cite{GC2007a} extended Watts' model to a finite fraction of initiators. Such extension appears to make a drastic impact on the cascade transition as a function of the average degree $z$, even making the transition to be discontinuous for relatively small values of $z$. Further work along this line has generalized the study to degree-correlated~\cite{JPG2008,DP2009}, directed~\cite{GK2010}, weighted~\cite{HG2013}, small-world~\cite{CEM2007}, modular~\cite{GC2007b}, clustered~\cite{IHN2010,HMG2011}, temporal~\cite{KH2013,BSP2014}, and multiplex~\cite{BLG2012,YG2012,LBG2014} networks, etc. On the other hand, studies have been carried out on contagion mechanisms. Dodds and Watts~\cite{DW2004,DW2005} proposed a generalized contagion model incorporating individual memory, variable magnitude of exposure, and susceptibility heterogeneity. They identified three basic classes of contagion models based on the memory length and the probabilities of being infected by one and two exposures, respectively. Research has also been conducted on more complicated threshold models. For example, one interesting work~\cite{FJP2011} introduced syngergistic effects from nearby neighbours, which can be mapped into a dynamical bond percolation with spatial correlations. Another study~\cite{NM2013} decomposed the motivation for a node to adopt a new behaviour as a combination of personal preference, the average of the states of each node's neighbours and the system average. It is worth mentioning that
Melnik et al.~\cite{SM2013} considered the threshold model with multi-stages and found that global cascades can be driven not only by high-stage influencers but also by low-stage ones. More recently, Wang et al.~\cite{WW2015a,WW2015b} proposed a contagion model with reinforcement derived from nonredundant information memory. They used a spreading threshold model as a specific example to understand the memory effect. Ruan et al.~\cite{RZY2015} considered individual conservativeness and studied Watts' model with mechanisms of spontaneous adoption and complete reluctance to adoption.

Most of these previous studies have assumed that whether or not a node gets activated depends on the states of its neighbours, ignoring the asymmetry of social interactions. During the diffusion of an entity or influence among individuals, however, transmission depends on both the probability of \emph{giving} it and the probability of \emph{catching} it, which are usually distinct. Although the adoption threshold, which represents the catching probability, has stimulated a rapid acceleration of research work, little attention has been paid to the giving probability. In reality, an individual surrounded by many active neighbours is easier to become a supporter of the issue and has a tendency towards convincing others that are insensitive~\cite{CD2011,ZZL2013,SMP2014}. In accordance with the catching dynamics, we introduce a persuasion threshold to the giving dynamics; that is, an activated individual can convince her inactivate friends if the active fraction among her friends is larger than a critical fraction. Based on this, we propose a ($\phi,\phi'$)-threshold model, where $\phi$ and $\phi'$ denote the adoption and persuasion thresholds, respectively. We study analytically and numerically the model on Poisson and power-law networks. As found earlier in the $\phi$-threshold model~\cite{DJW2002,GC2007a,PS2013}, a global cascade is not triggered when the average degree $z$ of nodes is either too low or too high. However, large cascades are realized within an intermediate range of $z$, which is referred to as the cascade window. When the persuasion threshold is taken into consideration, the cascade window becomes wider and the low-degree transition may be discontinuous in certain parameter regimes. We also investigate the impact of heterogeneous degrees and thresholds on the system dynamics, revealing that both of the heterogeneities make the dynamics much richer because of the ambiguous role of $\phi'$. To the best of our knowledge, such dynamics could not be observed in single threshold models.

\section*{Results}

\begin{figure}[ht]
\centering
\includegraphics[width=1.5\linewidth]{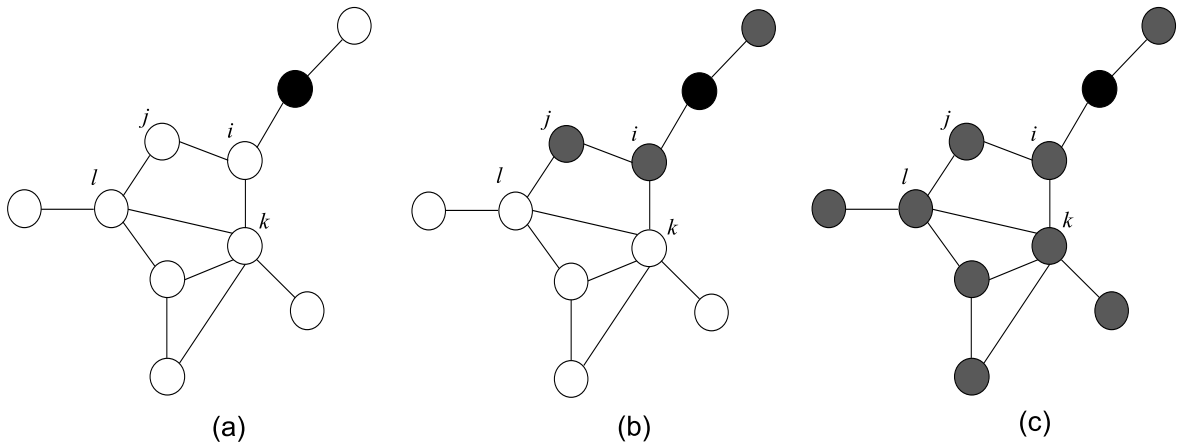}
\caption{\textbf{Illustration of cascade clusters for $\phi$- and $(\phi, \phi')$-threshold models.} (a) The initial network has one active seed and the rest nodes are inactive. The adoption and persuasion thresholds are $\phi=0.3$ and $\phi'=0.4$, respectively. (b) The final activated nodes for the $\phi$-threshold model. Since the activated fractions in the neighbourhoods of nodes $k$ and $l$ are $1/5$ and $1/4$, respectively, which are less than the adoption threshold $\phi=0.3$, both of them are not activated. (c) The final activated nodes for the $(\phi, \phi')$-threshold model. Although the adoption rule is unsatisfied for $k$ and $l$, the activated fractions in the neighbourhoods of $i$ and $j$ are $2/3$ and $1/2$ (see Fig. 1(b)), respectively, both of which are larger than the persuasion threshold $0.4$. According to the persuasion rule, $i$ and $j$ can activate $k$ and $l$, respectively.}
\label{fig:1}
\end{figure}

In the ($\phi,\phi'$)-threshold model, each node of a network can be in one of the two discrete states: inactive and active. Initially a fraction $\rho_0$ of nodes are chosen randomly from the network to be active, and the others are inactive. At each time step, an inactive node $i$ will be activated  in either of the following two cases: i) the active fraction of the neighbours of node $i$ is larger than its adoption threshold $\phi_i$, which is defined as the adoption dynamics; or ii) the adoption dynamics does not occur, but there is at least one active neighbour $j$ of the node $i$ being a {\it persuader}, i.e., the active fraction in the neighbourhood of $j$ is larger than the persuasion threshold $\phi'_j$. We call it the persuasion dynamics. Once a node is activated, it remains unchanged. The system evolves according to above rules until no further activation occurs. Both $\phi$ and $\phi'$ are random variables drawn from probability distribution functions $f(\phi)$ and $g(\phi')$ with $\int_{0}^{1}f(\phi)d\phi=1$ and $\int_{0}^{1}g(\phi')d\phi'=1$, respectively. In the context of innovation diffusion or opinion spreading, the ($\phi,\phi'$)-threshold model can be described as follows. Consider a population of users to adopt a new product or idea, a small fraction of users are the initial spreaders. For a user in the rest part of population, if the adopted fraction of her friends is larger than her adoption threshold, she will use the product. Meanwhile, a friend of the user with many adopted neighbours may tend to be more committed and thus has a good chance to convince her. As shown in Fig.~\ref{fig:1}, the activated sizes are different with or without persuasion.

\textbf{The impact of persuasion.} The persuasion threshold $\phi'$ represents the situation that a persuader activates her inactive friends. Therefore it gives rise to global cascades. According to the model definition, the higher $\phi'$, the lower the persuasion possibility is. In the extreme case $\phi'=1$, our model reduces to Watts' model~\cite{DJW2002}. In this scenario, the cascade condition in random networks for one seed is $\sum_{k}k(k-1)\rho(k)P(k)=z$, where $\rho(k)$ represents the distribution of vulnerable nodes and $P(k)$ is the distribution of all the nodes. While the network is sparse, the criterion of the cascade is $\phi<1/z$. But if the number of initiators is sufficiently large, large cascades will occur despite of $\phi$.

\begin{figure}[ht]
\centering
\includegraphics[width=1.5\linewidth]{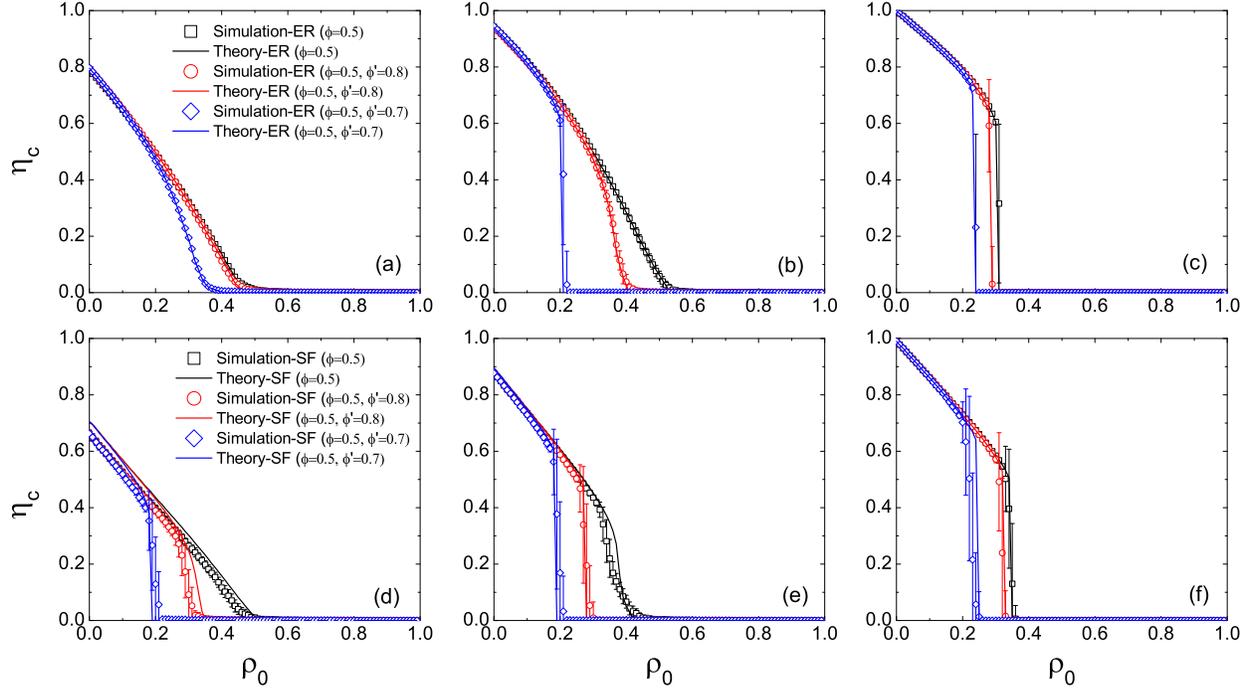}
\caption{\textbf{Normalized size of the giant component of inactive nodes $\eta_c$ in the stable states as a function of the seed fraction $\rho_0$.} Both the adoption and persuasion thresholds are uniform. Symbols represent simulation results on ER (upper panel) and SF (lower panel) networks of $N=10^4$ nodes and average degree $z=2$ (left column), $3$ (middle column) and $10$ (right column), respectively. All the results are averaged over $50$ realizations of the model, each of which is performed on $50$ network configurations. Error bars are the standard deviations of the means. Solid lines are theoretical predictions by Eq.~(\ref{goodcore}).}
\label{fig:2}
\end{figure}

In Fig.~\ref{fig:2} we plot the normalized size of the final giant component of inactive nodes $\eta_c$ as a function of the seed fraction $\rho_0$. Both the adoption and persuasion thresholds are uniform. Simulations are performed on Erd\H{o}s-R\'{e}nyi (ER)~\cite{ER1959} and scale-free (SF) networks~\cite{MC2005}, respectively. The adoption threshold is fixed at $\phi=0.5$, and two different values of the persuasion threshold $\phi'=0.7$ and $0.8$ are considered, representing different extents of persuasion respectively. From left to right, the values of the average degree of nodes are $z=2$, $3$, and $10$, respectively. All the symbols correspond to numerical results of the model. The solid lines represent analytical results based on Eq.~(\ref{goodcore}) in the methods section. For ER networks, one notices smooth curves separating two phases for $z=2$ (see Fig.~\ref{fig:2}(a)), which defines the transition point $\rho_c$. Global cascades are observed when $\rho_0>\rho_c$. As $z$ increases to $3$ (see Fig.~\ref{fig:2}(b)), different behaviours can be observed: the curve is still smooth without persuading effect for $\phi=0.5$. While the persuasion is considered and of a high value ($\phi'=0.7$), the curve drops abruptly from a finite size to zero at $\rho_c$, indicating a discontinuous transition. For $z=10$ (see Fig.~\ref{fig:2}(c)), the system exhibits discontinuous transitions for all the parameter values. Thus, the addition of the persuasion mechanism not only causes $\rho_c$ to be smaller but also changes transition behaviour. These conclusions hold in SF networks as well (see the lower panel of Fig.~\ref{fig:2}). We also studied $\eta_c$ for $\phi=0.6$ and obtained similar phenomena (see Supplementary Fig. S1 online).

\begin{figure}[ht]
\centering
\includegraphics[width=0.8\linewidth]{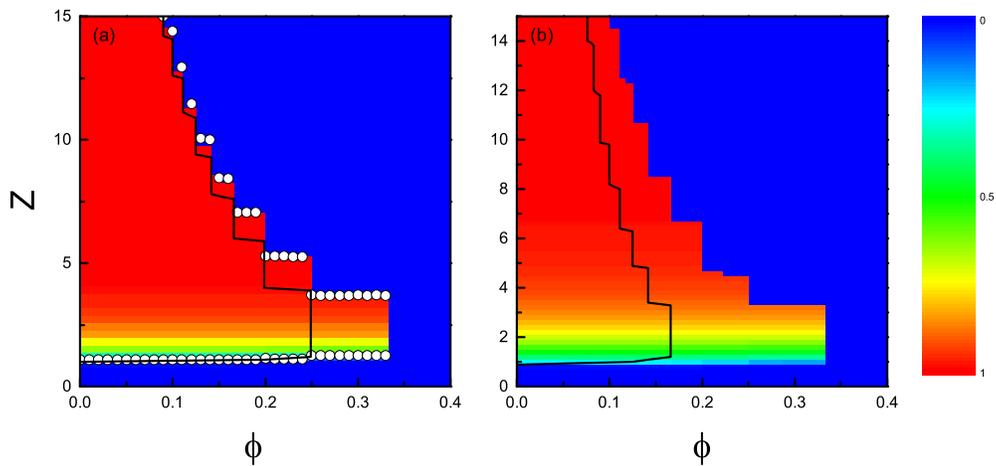}
\caption{\textbf{Effect of the persuasion threshold on the cascade window in ER (a) and SF (b) networks.} Color-coded values represent analytical results of the final fraction of active nodes $\rho$ based on Eq.~(\ref{eta}) with seed fraction $\rho_0=10^{-4}$. The persuasion threshold is $\phi'=0.5$. Circles correspond to simulation results of the present model. Solid lines are cascade boundaries of Watts' model on the ER and SF networks, respectively.}
\label{fig:3}
\end{figure}

Next, we investigate the persuading effect on the cascade window which delineates the region where global cascades can occur. We carry out simulations on ER and SF networks for the uniform threshold with seed fraction $\rho_0=10^{-4}$. In Fig.~\ref{fig:3} the color-coded values are analytical results of the final fraction of activated nodes $\rho$ ($=1-\eta$) for $\phi'=0.5$, where $\eta$ is the stable fraction of inactive nodes (see Eq.~(\ref{eta}) in the methods section). For a given $\phi$, $\rho$ becomes smaller as $z$ decreases, since the network connectivity is low which limits cascade propagation. For the ER networks the cascade boundaries are in good agreement with numerical results (open circles). While for the SF networks the fluctuation of the numerical results occurs (not shown) because of the degree heterogeneity. The solid lines in Figs.~\ref{fig:3}(a) and ~\ref{fig:3}(b) correspond to the results of Watts' model on the ER and SF networks, respectively. One notices that the persuasion mechanism makes both networks more vulnerable. However, the influence is weakened as $z$ increases, since the possibility for a node being a persuader is low when it is surrounded by many inactive neighbors. We also calculated $\rho$ for $\rho_0=10^{-3}$. Again, similar phenomena can be observed (see Supplementary Fig. S2 online).

\begin{figure}[ht]
\centering
\includegraphics[width=1.5\linewidth]{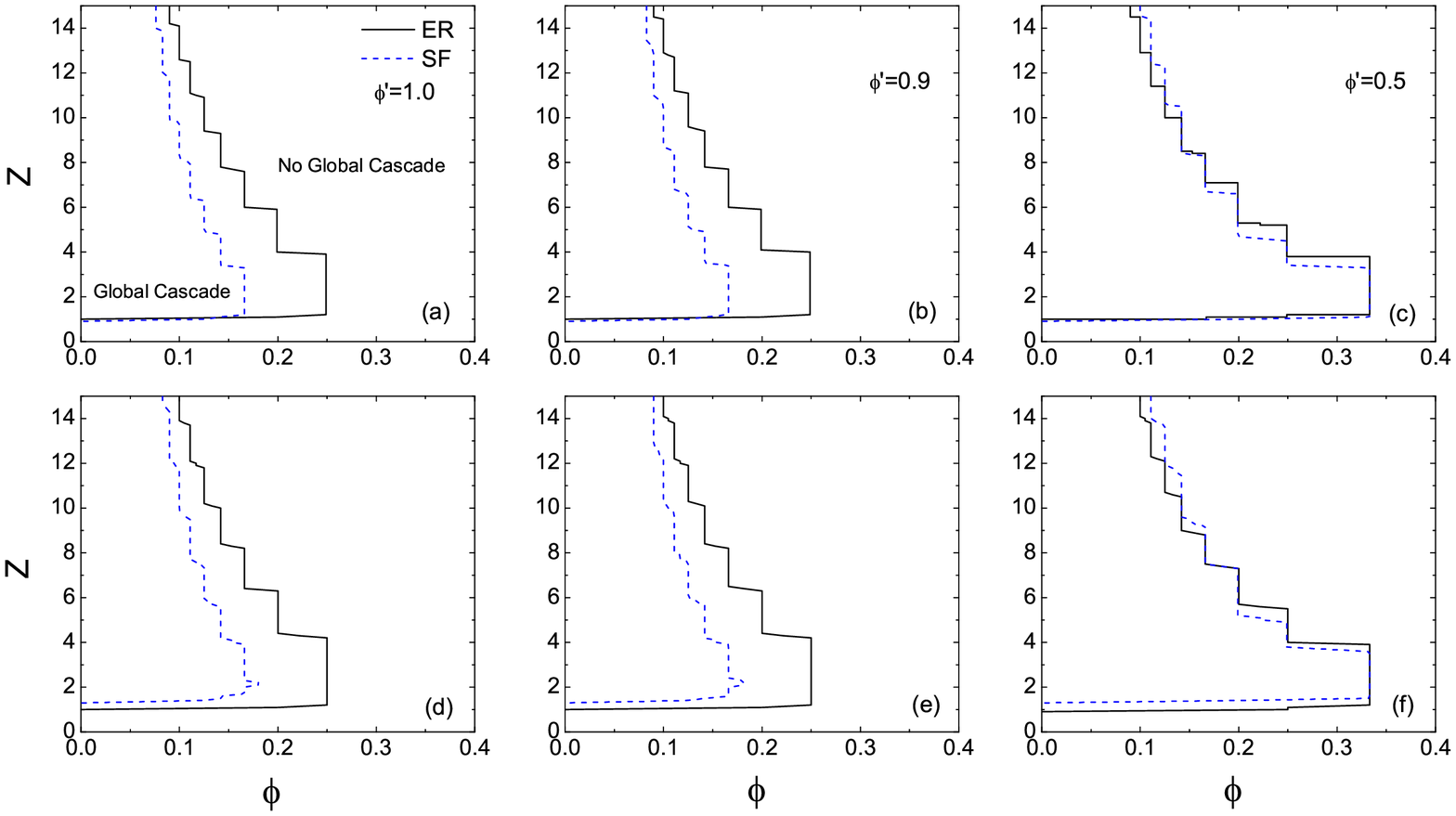}
\caption{\textbf{Comparison of the systemic stability of ER and SF networks.} Both the adoption and persuasion thresholds are uniform. The seed fractions are $N=10^{-4}$ (upper panel) and $10^{-3}$ (lower panel), respectively. The values of the persuasion threshold are $\phi'=1.0$ (left column), $0.9$ (middle column) and $0.5$ (right column), respectively.}
\label{fig:4}
\end{figure}

\textbf{The impact of heterogeneity.} Although SF networks show qualitatively the same behaviour as those of ER networks with the persuasion threshold, i.e, the decrease of $\phi'$ results in a larger cascade window. However, they are quantitatively different. We compare cascade windows for both networks in Fig.~\ref{fig:4}, where the upper and lower panels correspond to $\rho_0=10^{-4}$ and $10^{-3}$, respectively. Under weak persuading effect ($\phi'=0.9$), the cascade windows in the SF networks are smaller than those in the ER networks (see Figs.~\ref{fig:4}(b) and \ref{fig:4}(e)), which is similar to the $\phi$-threshold model (see Figs.~\ref{fig:4}(a) and \ref{fig:4}(d)), implying that heterogeneous networks are more robust against random shocks than homogeneous networks with the same connectivity. While the persuading effect is enhanced ($\phi'=0.5$), the SF networks are much more impacted than the ER networks: when the connectivity is sufficiently sparse ($z<10$ in Fig.~\ref{fig:4}(c) and $z<9$ in Fig.~\ref{fig:4}(e)), the systemic stability of the SF networks is still better than that of the corresponding ER networks with the same connectivity; but when it is sufficiently dense ($z>10$ in Fig.~\ref{fig:4}(c) and $z>9$ in Fig.~\ref{fig:4}(e)), the SF networks tend to become less stable than the ER networks. We conclude that heterogeneous networks are more vulnerable to perturbations when the persuading effect is strong enough.

\begin{figure}[ht]
\centering
\includegraphics[width=1.2\linewidth]{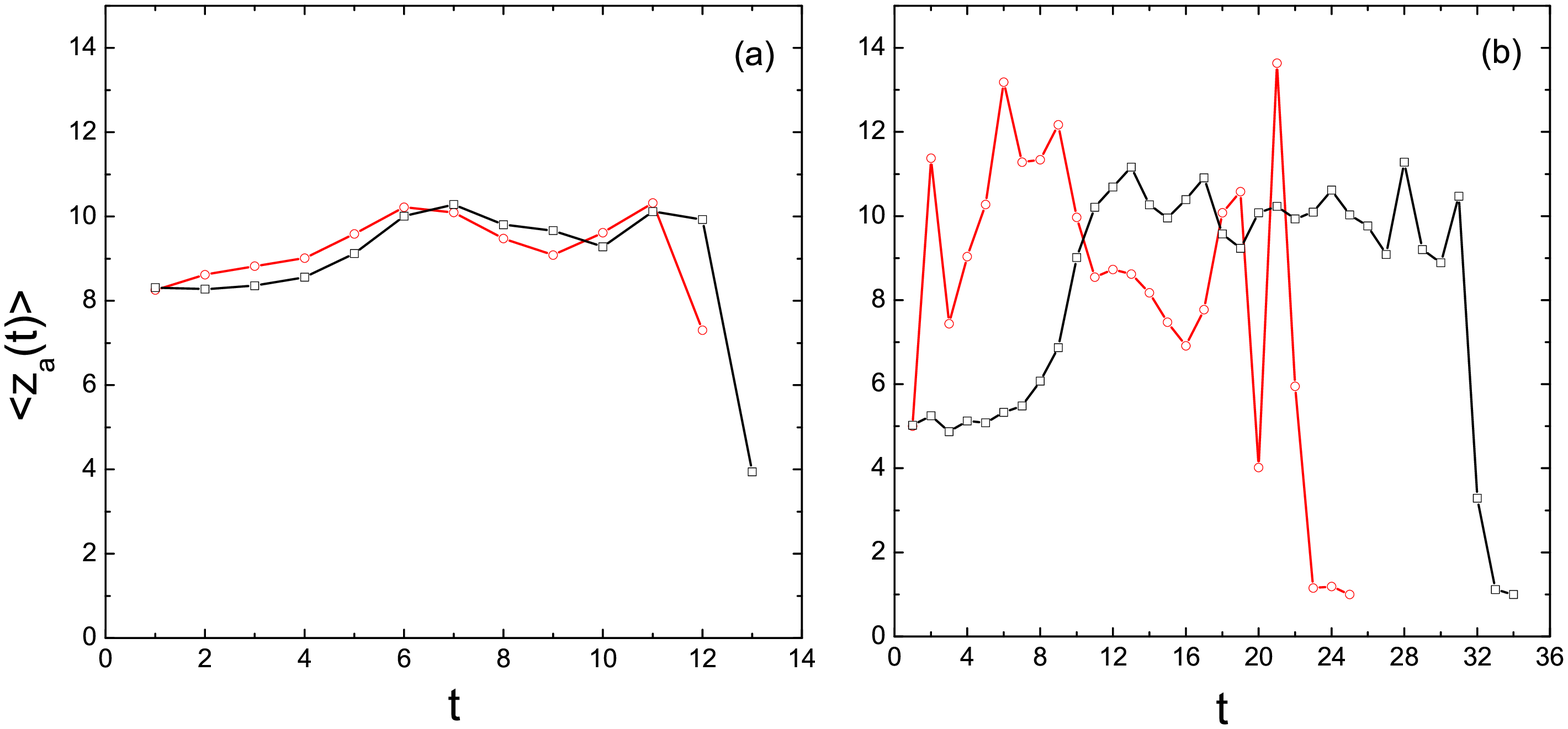}
\caption{\textbf{Temporal plots of the average degree of newly activated nodes for ER (a) and SF (b) networks.}
The seed fraction is $\rho_0=10^{-4}$ and the adoption threshold is $\phi=0.1$. Squares and circles correspond to simulation results of the model without and with persuasion, respectively.}
\label{fig:5}
\end{figure}

To get a clear inspection on this point, we have calculated average degrees of newly activated nodes in ER and SF networks with $z=10$. The values of the persuasion threshold are $\phi'=1.0$ (without persuading effect) and $0.5$ (with persuading effect), respectively. For the ER network (see Fig.~\ref{fig:5}(a)), both plots display the same trend, implying that the persuading effect does not change the cascade order of the network. Therefore, low- and average-degree nodes are still responsible for triggering large cascades~\cite{DJW2002}. By contrast, there is a big difference for the SF network with and without persuasion (see Fig.~\ref{fig:5}(b)). While the persuasion dynamics is considered, the cascade invades high-degree nodes quickly and then spans the whole network.

\begin{figure}[ht]
\centering
\includegraphics[width=1\linewidth]{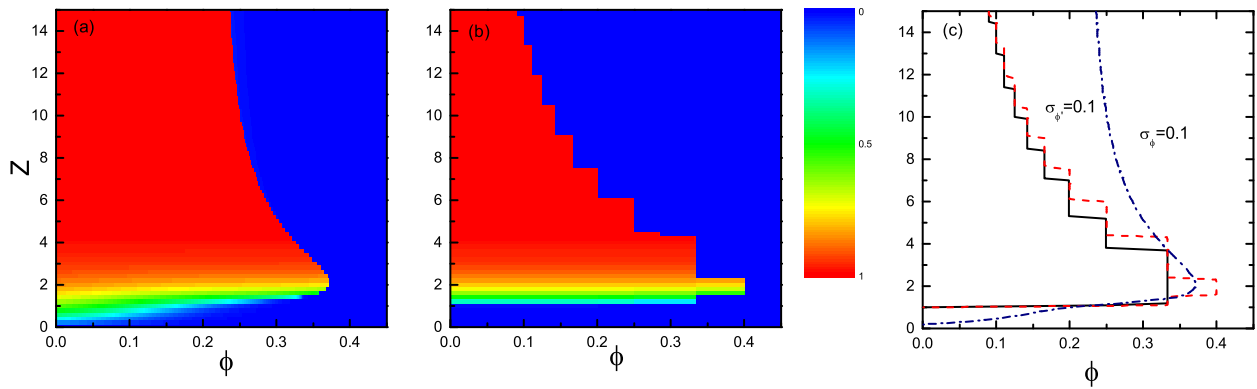}
\caption{\textbf{Effect of heterogeneous thresholds on the cascade window in ER networks.}
Cascade windows on the ($\phi$, $z$) plane with seed fraction $\rho_0=10^{-4}$ in the ER networks where the adoption threshold is normally distributed with a mean $\phi$ and a standard deviation $\sigma=0.1$ (a) and the persuasion threshold is normally distributed with a mean $\phi'=0.5$ and a standard deviation $\sigma=0.1$ (b), respectively. The color codes represents analytical predictions of the final fraction of active nodes $\rho$ based on Eq.~(\ref{eta}). (c) Comparison of the cascade windows for both thresholds. The solid line corresponds to the result of Watts' model with the uniform adoption threshold.}
\label{fig:6}
\end{figure}

We now turn to the effect of the threshold heterogeneity on the cascade dynamics. Figures~\ref{fig:6}(a) and ~\ref{fig:6}(b) show cascade windows for ER networks with the adoption and persuasion thresholds respectively following the Gaussian distribution. The standard deviation is $\sigma=0.1$ representing fluctuations. For both cases, the active fraction increases with the network connectivity for a given $\phi$ inside the cascade window. Figures~\ref{fig:6}(c) shows the comparison of the cascade windows. It is clear that the threshold heterogeneity appears to increase the likelihood of global cascades. Especially for networks with dense connectivity, the adoption threshold has an overwhelming influence compared to that of the persuasion threshold. Such difference can be understood: when the persuasion threshold $\phi'$ is Gaussian distributed, there is a higher probability for the active fraction of a adopter's neighbors exceeds the persuasion threshold than the uniform case, which results in a stronger inducing effect on cascade propagation. While the network connectivity increases, an adopter is surrounded by many inactive neighbors, which leads to a lower chance for her to be a persuader. Hence the number of persuaders is reduced, leading to a less significant inducing effect. On the contrary, when the adoption threshold $\phi$ follows the Gaussian distribution, the increase of the network connectivity brings about more early adopters with lower adoption thresholds and relatively higher degrees, which accelerates cascade propagation. This conclusion is valid for a wider range of the seed fraction (see Supplementary Fig. S3 online).

\begin{figure}[ht]
\centering
\includegraphics[width=1.5\linewidth]{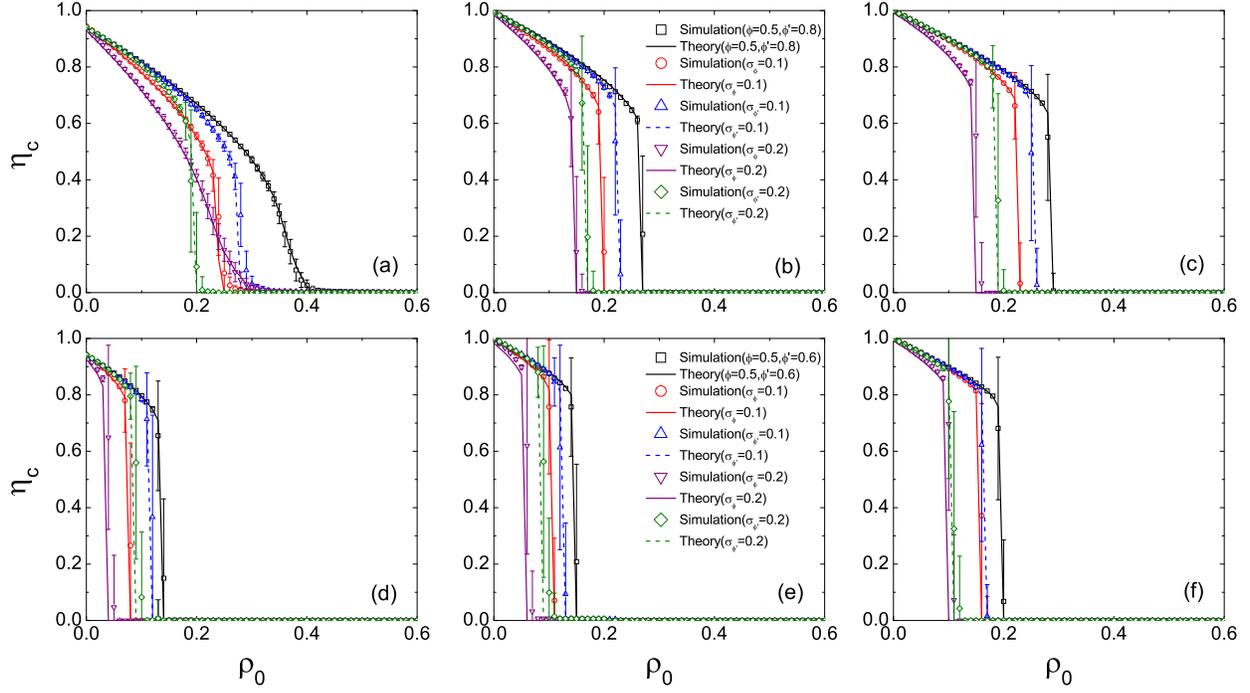}
\caption{\textbf{Effect of heterogeneous thresholds on the size of the giant component of inactive nodes $\eta_c$ in ER networks.} Symbols represent simulation results of the ER networks of size $N=10^4$ and an average degree $z=3$ (left column), $5$ (middle column) and $10$ (right column), respectively. The values of the persuasion threshold are $\phi'=0.8$ (upper panel) and $0.6$ (lower panel), respectively. Error bars are the standard deviations of the means. Lines are theoretical predictions by Eq.~(\ref{goodcore}).}
\label{fig:7}
\end{figure}

Finally, we examine the threshold heterogeneity on the transition behaviour. Figure~\ref{fig:7} shows the normalized size of the giant component of inactive nodes $\eta_c$ as a function of the seed fraction $\rho_0$ in ER networks. The upper and lower panels correspond to $\phi'=0.8$ and $0.6$, respectively. When the network connectivity is sufficiently sparse ($z=3$) and the persuading effect is sufficiently weak ($\phi'=0.8$), the system exhibits the continuous transition for $\sigma=0$ (see Fig.~\ref{fig:7}(a)). When the adoption heterogeneity is allowed but of a relatively small value ($\sigma_{\phi}=0.1$), the transition becomes discontinuous. For larger adoption heterogeneity ($\sigma_{\phi}=0.2$) the system exhibits the continuous transition again. This is qualitatively the same as the $\phi$-threshold model~\cite{PDK2015}. On the contrary, the increase of the persuasion heterogeneity $\sigma_{\phi'}$ results in the monotonic change in the phase transition: the transition is always discontinuous in dense networks or in networks with strong persuasion (see Figs.~\ref{fig:7}(b)-\ref{fig:7}(f)).

\section*{Discussion}

In spite of its simplicity, the threshold model has attracted much attention from social, mathematical, physical, and biological communities. Although there is a rapid acceleration of work on this topic, very few studies have considered the asymmetry of social interactions. In this paper, we decomposed the spread of an entity or influence as a combination of giving and catching operations, with chances of happening ruled by persuasion and adoption probabilities, respectively. Based on this, we proposed the $(\phi,\phi')$-threshold model.

The focus of the present work is to identify the effects of the adoption and persuasion thresholds. We first studied the stable giant component of inactive nodes as a function of initiators for uniform thresholds. The introduction of the persuasion threshold not only facilitates global cascades, but also may turn the transition from continuous to discontinuous, even when the network connectivity is low. This local mechanism is expected to explain abrupt breakdown of real systems. We also explored the effect of the persuasion threshold on the cascade window. When the persuading effect is weak, heterogeneous networks are more robust against random shocks than homogeneous networks; when the persuading effect is strong, however, homogeneous networks are more stable. Thus, the uniform persuasion threshold has stronger effects on the transmission dynamics than the uniform adoption threshold. Finally, we studied the effect of threshold heterogeneity on systemic stability. Although both heterogeneities give rise to the cascade, the adoption heterogeneity has an overwhelmingly stronger influence than the persuasion heterogeneity for the networks that are sufficiently dense: the higher the network connectivity is, the less significant effects the persuasion threshold has on the cascade dynamics, since the local stability of nodes suppresses the persuading effect. These striking results indicate that the persuasion dynamics is important for understanding contagion processes on social networks with asymmetry. It is interesting to incorporate more structural characteristics in asymmetric contagion models, such as direction, weights, and correlations.

\section*{Methods}

Given an uncorrelated network of $N$ nodes following the degree distribution $P(k)$, a fraction $\rho_0$ of nodes are chosen randomly to be active. According to the model definition, we obtain the stable fraction of inactive nodes:
\begin{equation}
\eta=(1-\rho_0)\sum_{k_{i}=0}^{k_{max}}P(k_{i})
\sum_{s=0}^{k_{i}}C_{k_i}^{s}(1-\alpha-\beta)^s\beta^{k_{i}-s}F\left(\frac{k_i-s}{k_i}\right),\label{eta}
\end{equation}
where $F(x)$ denotes the probability that the adoption threshold $\phi$ of a node is no less than $x$. $\alpha$ represents the probability that a random neighbor $j$ of the inactive node $i$ is active and the active fraction in the neighbourhood of $j$ is larger than the persuasion threshold $\phi'_j$. $\beta$ represents the probability that a random neighbor $j$ of the inactive node $i$ is active and the active fraction in the neighbourhood of $j$ is less than the persuasion threshold $\phi'_j$. Following the ideas of Refs.~\cite{GC2007a,ZZL2013}, we obtain the self-consistent equations for the two probabilities:
\begin{eqnarray}
\alpha&=&\rho_0\sum_{k_j=0}^{k_{max}}Q(k_j)\sum_{s=0}^{k_j}C_{k_j}^{s}(1-\delta)^{s}\delta^{k_{j}-s}\left[1-G\left(\frac{k_j-s}
            {k_{j}+1}\right)\right]\nonumber\\
      &&+(1-\rho_0)\sum_{k_{j}=0}^{k_{max}}Q(k_{j})\sum_{s=0}^{k_j}C_{k_{j}}^{s}(\alpha+\beta)^{k_{j}-s}\left[1-F\left(\frac{k_{j}-s}{k_{j}+1}
            \right)\right]\nonumber\\
      &&\times\sum_{m=0}^{s}C_{s}^{m}(1-\alpha-\beta-\gamma)^{m}\gamma^{s-m}\left[1-G\left(\frac{k_{j}-m}{k_{j}+1}\right)\right]\nonumber\\
      &&+(1-\rho_0)\sum_{k_{j}=1}^{k_{max}}Q(k_{j})\sum\limits_{s=0}^{k_{j}-1}C_{k_{j}}^{s}\left[(\alpha+\beta)^{k_{j}-s}-\beta^{k_{j}-s}\right]F\left(\frac{k_{j}-s}{k_{j}+1}\right)\nonumber\\
      &&\times\sum_{m=0}^{s}C_{s}^{m}(1-\alpha-\beta-\gamma)^{m}\gamma^{s-m}\left[1-G\left(\frac{k_{j}-m}{k_{j}+1}\right)\right],\label{alpha}\\
\beta&=&1-\alpha-(1-\rho_0)\sum_{k_{j}=0}^{k_{max}}Q(k_{j})\sum_{s=0}^{k_{j}}C_{k_{j}}^{s}(1-\alpha-\beta)^{s}\beta^{k_{j}1-s}F\left(\frac{k_{j}-s}{k_{j}+1}\right),\label{beta}
\end{eqnarray}
where $\gamma$ refers to the critical case separating $\alpha$ and $\beta$, and $\delta$ describes the probability that a random neighbor $j$ of the active node $i$ is active, written as
\begin{eqnarray}
\gamma&=&(1-\rho_0)\sum_{k_{j}=0}^{k_{max}}Q(k_{j})\sum_{s=0}^{k_{j}}C_{k_{j}}^{s}(1-\alpha-\beta)^{s}\beta^{k_{j}-s}\left[F\left(
            \frac{k_{j}-s}{k_{j}}\right)-F\left(\frac{k_{j}+1-s}{k_{j}+1}\right)\right],\label{gamma}\\
\delta&=&1-(1-\rho_0)\sum_{k_{j}=0}^{k_{max}}Q(k_{j})\sum\limits_{s=0}^{k_{j}}C_{k_{j}}^{s}(1-\alpha-\beta)^{s}\beta^{k_{j}-s}F\left(\frac{k_{j}+1-s}
            {k_{j}+1}\right).\label{delta}
\end{eqnarray}
$G(x)$ denotes the probability that the persuasion threshold $\phi'$ of a node is no less than $x$. $Q(k)\equiv (k+1)P(k+1)/z$ is the excess degree distribution. One can solve the above equations using a simple iterative scheme, and finally get the stable size of the giant component of the inactive nodes:
\begin{equation}
\eta_c=(1-\rho_0)\sum_{k_{i}=1}^{k_{max}}P(k_{i})\sum_{s=1}^{k_{i}}C_{k_{i}}^{s}\beta^{k_{i}-s}F\left(\frac{s}{k_{i}}\right)
        \sum_{m=1}^{s}C_{s}^{m}(1-\alpha-\beta-\theta)^{m}\theta^{s-m},\label{goodcore}
\end{equation}
where $\theta$ is the probability that a random neighbor $j$ of the inactive node $i$ is inactive but not belonging to the giant component of the inactive nodes, given by
\begin{equation}
\theta=(1-\rho_0)\sum_{k_j=0}^{k_{max}}Q(k_j)\sum_{s=0}^{k_j}C_{k_j}^{s}\theta^{s}\beta^{k_j-s}F\left(\frac{k_j-s}{k_j+1}\right).
\end{equation}

\section*{Acknowledgements}

The authours thank Gaoxi Xiao and Ming Tang for helpful discussions, and are grateful to referees for valuable comments. This work was supported by NSFC grant 11331009 and STCSM grant 13ZR1416800.

\section*{Author contributions statement}

L.-J.Z., X.-J.X. and X.F. conceived the study; W.-M.H., L.-J.Z. and X-J.X. implemented analytical calculation and numerical simulation of the model; W.-M.H., L.-J.Z., X.-J.X. and X.F. wrote the paper.

\section*{Additional information}

\textbf{Supplementary information} accompanies this paper at http://www.nature.com/srep; \textbf{Competing financial interests:} The authors declare no competing financial interests.

\end{document}